\documentclass[12pt,a4paper]{article}
\usepackage{authblk}
\usepackage{amsmath,amssymb}
\usepackage{graphicx,psfrag,color}
\usepackage{epsfig}
\usepackage{cite}

\newcommand{\im}{\mathrm i}

\newcommand{\lgn}{\operatorname{ln}}

\newcommand{\eq}{\begin{equation}}
\newcommand{\en}{\end{equation}}
\newcommand{\bear}{\begin{eqnarray}}
\newcommand{\ear}{\end{eqnarray}}

\title{Magnetocaloric effect in the spin-$1/2$ chain with competing interactions}

\author{T.S. Tavares$^{(2)}$ and G.A.P. Ribeiro$^{(1,2)}$\footnote{pavan@df.ufscar.br}  \\ $^{(1)}$  C.N. Yang Institute for Theoretical Physics, \\
State University of New York at Stony Brook, \\
Stony Brook, NY 11794-3840, USA }
\affil{$^{(2)}$ Departamento de F\'{i}sica, Universidade Federal de S\~ao Carlos \\ 13565-905 S\~ao Carlos-SP, Brazil}

\begin{document}
\maketitle

\begin{abstract}
We study the magnetocaloric effect for the integrable antiferromagnetic Heisenberg spin chain with competing interactions. We computed  the Gr\"uneisen parameter, which is closely related to the magnetocaloric effect, for the quantum spin chain in the thermodynamical limit. This is obtained by means of the solution of a set of non-linear integral equations, which describes the thermodynamics of the model. We also provide results for the entropy $S$ and the isentropes in the $(H,T)$ plane. 
\end{abstract}
%\centerline{Keywords: Magnetocaloric effect, integrable systems}
\thispagestyle{empty}

\newpage

\section{Introduction}

The magnetocaloric effect refers to any temperature change of magnetic systems induced by an adiabatic 
variation of the magnetic field. In the last decade, this effect received increasing attention from experimental as well as theoretical viewpoint \cite{TISHIN,TSOKOL}. 

Moreover, the magnetocaloric effect has been described by means of exact and numerical tools in various one-dimensional interacting spin systems\cite{HONECKER2004,HONECKER2009,BOSTREM,ROSCH2003,ROSCH2005,TRIPPE,RIBEIRO,TOPILKO,GALISOVA}. Besides its quantitative description, it has been also observed that there is an enhancement of the magnetocaloric effect in the vicinity of quantum critical points\cite{ROSCH2003,ROSCH2005}.

Recently these studies have been extended to the case of spin-$1/2$ XX chains with three-spin interactions \cite{TOPILKO} and Ising-Heisenberg four-spin interactions\cite{GALISOVA}. A natural extension of these results would be the study of integrable spin-$1/2$ Heisenberg model with additional interaction among spins further apart. This is specially interesting due to the existence of quasi one-dimensional quantum magnets, which can be described by spin-$1/2$ with next-nearest neighbour interactions\cite{EXP}. 

In this paper, we are interested in the exact computation of the magnetocaloric effect for the integrable antiferromagnetic spin-$1/2$ Heisenberg model with additional interactions among three and four spins. This can be obtained from the related quantity called Gr\"uneisen parameter $\Gamma_H=\frac{1}{T}\left(\frac{\partial T}{\partial H}\right)_S=-\frac{1}{C_H}\left(\frac{\partial M}{\partial T} \right)_H$. We also compute the entropy and the isentropes in the $(H,T)$ plane. These integrable spin chain with competing interactions (three-spin, four-spin and so on) can be constructed by an arbitrary number of staggering parameters \cite{ZVYAGINstg0,ZVYAGINstg,SEDRAKYAN2000,SEDRAKYAN2003}.

The thermodynamic quantities required to achieve this goal, like entropy, specific heat and magnetization, are determined as a function of temperature and magnetic field. This is obtained by means of the solution of a finite set of NLIE which arises from the QTM approach \cite{ZVYAGIN,KLUMPER03,TRIPPE0,THIAGO}.

This paper is organized as follows. In section \ref{integra}, we outline the integrable Hamiltonians and the associated integral equations. In section \ref{gruneisen}, we present our results for the magnetocaloric effect and the isentropes in the $(H,T)$ plane. Our conclusions are given in
section \ref{conclusion}.

\section{Hamiltonian and integral equations}\label{integra}

The integrable spin-$1/2$ Heisenberg model with interaction among more spins can be constructed within the quantum inverse scattering method by the introduction of certain staggering spectral parameter\cite{ZVYAGINstg0,ZVYAGINstg,SEDRAKYAN2000,SEDRAKYAN2003}. We give below the explicit expression for the three and four spin interaction Hamiltonians \cite{THIAGO},
\begin{equation}
{\cal H}_3= \frac{1}{8(1+  \theta_1^2)}\sum_{i=1}^{2 L}[-(2+ \theta_1^2)+ 2 \vec{\sigma}_{i} \cdot \vec{\sigma}_{i+1}+ \theta_1^2 \vec{\sigma}_{i} \cdot \vec{\sigma}_{i+2}+(-1)^i \theta_1 \vec{\sigma}_{i} \cdot \vec{\sigma}_{i+1}\times \vec{\sigma}_{i+2}],
\label{HM3}
\end{equation}
and
\begin{multline}
{\cal H}_{4}=\frac{1}{2}\sum_{i=1}^{3 L} c_{0, i}(\theta_1,\theta_2)+c_{1,i}(\theta_1,\theta_2) \vec{\sigma}_{i} \cdot \vec{\sigma}_{i+1}+c_{2,i}(\theta_1,\theta_2)\vec{\sigma}_{i} \cdot \vec{\sigma}_{i+2}+\\ c_{3,i}(\theta_1,\theta_2)\vec{\sigma}_{i} \cdot \vec{\sigma}_{i+1} \times \vec{\sigma}_{i+2}+c_{4,i}(\theta_1,\theta_2) \vec{\sigma}_{i} \cdot \vec{\sigma}_{i+3}+ c_{5,i}(\theta_1,\theta_2) \vec{\sigma}_{i} \cdot \vec{\sigma}_{i+1} \times \vec{\sigma}_{i+3}+\\ c_{6,i}(\theta_1,\theta_2)\vec{\sigma}_{i} \cdot \vec{\sigma}_{i+2}\times \vec{\sigma}_{i+3}+c_{7,i}(\theta_1,\theta_2) \vec{\sigma}_{i} \cdot(\vec{\sigma}_{i+1} \times(\vec{\sigma}_{i+2}\times \vec{\sigma}_{i+3})),
\label{HM4}
\end{multline}
where $\theta_i$ are free parameters. The coefficients $c_{k,i}$ have the periodicity property $c_{k,i+3}=c_{k,i}$ and their explicit forms are given in appendix A. These Hamiltonians can be viewed as a two-chain spin model with zigzag interchain interaction \cite{ZVYAGIN} and as a three-chain with multiple interchain interactions\cite{THIAGO}. Note that $\theta_i\rightarrow 0$ recover the usual Heisenberg spin chain.

The free-energy in the thermodynamic limit for the above spin chains (\ref{HM3}) and (\ref{HM4}) are given as follows,
\eq
f=e_0-\frac{1}{(M-1)\beta}\sum_{j=1}^{M-1} \left(K*\lgn{B\bar{B}}\right)(-\theta_{j-1}),
\en
where the ground state energy is given by
\eq
e_0=-\frac{1}{2}\int_{-\infty}^{\infty} \frac{1}{1+{\rm e}^{|k|}} {\Bigg|\frac{\sum_{j=1}^{M-1} {\rm e}^{\im k \theta_{j-1}}}{(M-1)} \Bigg|}^2{\rm d}k.
\nonumber
\en
The auxiliary functions $B=b+1,\bar{B}=\bar{b}+1$ are solutions a set of non-linear integral equations \cite{THIAGO},
\bear
\lgn{b(x)} &=& \beta d_+(x) + (K*\lgn{B(x)})(x) -(K*\lgn{\bar{B}(x)})(x+\im), \\
\lgn{\bar{b}(x)} &=& \beta d_-(x) - (K*\lgn{B(x-\im)})(x) +(K*\lgn{\bar{B}(x)})(x),
\ear
where $\beta=1/T$, $d_{\pm}(x)=-\frac{1}{M-1}\sum_{j=1}^{M-1} \frac{\pi}{2\cosh(\pi (x+\theta_{j-1}))} \pm \frac{H}{2}$, $\theta_0=0$ and the symbol $*$
denotes the convolution $f*g(x)=\frac{1}{2 \pi}\int_{-\infty}^{\infty} f(x-y)g(y)dy$. The kernel is given by $K(x)=\int_{-\infty}^{\infty}\frac{e^{-|k|/2+\im k x}}{2 \cosh{\left[ k/2 \right]}} dk $.

\section{Gr\"uneisen parameter and entropy}
\label{gruneisen}

In this section we will present the results for the Gr\"uneisen parameter, which is closely related to the magnetocaloric effect. We will also show the results for the entropy and the isentropes in the $(H,T)$ plane. The results were obtained from the numerical solution of the above non-linear integral equations. Typically, these equation are solved by iteration and it allows for accurate results of the thermodynamical quantities at finite temperatures and magnetic fields.

\begin{figure}[t!]
\begin{minipage}{0.5\linewidth}
\begin{center}
\includegraphics[width=1\columnwidth]{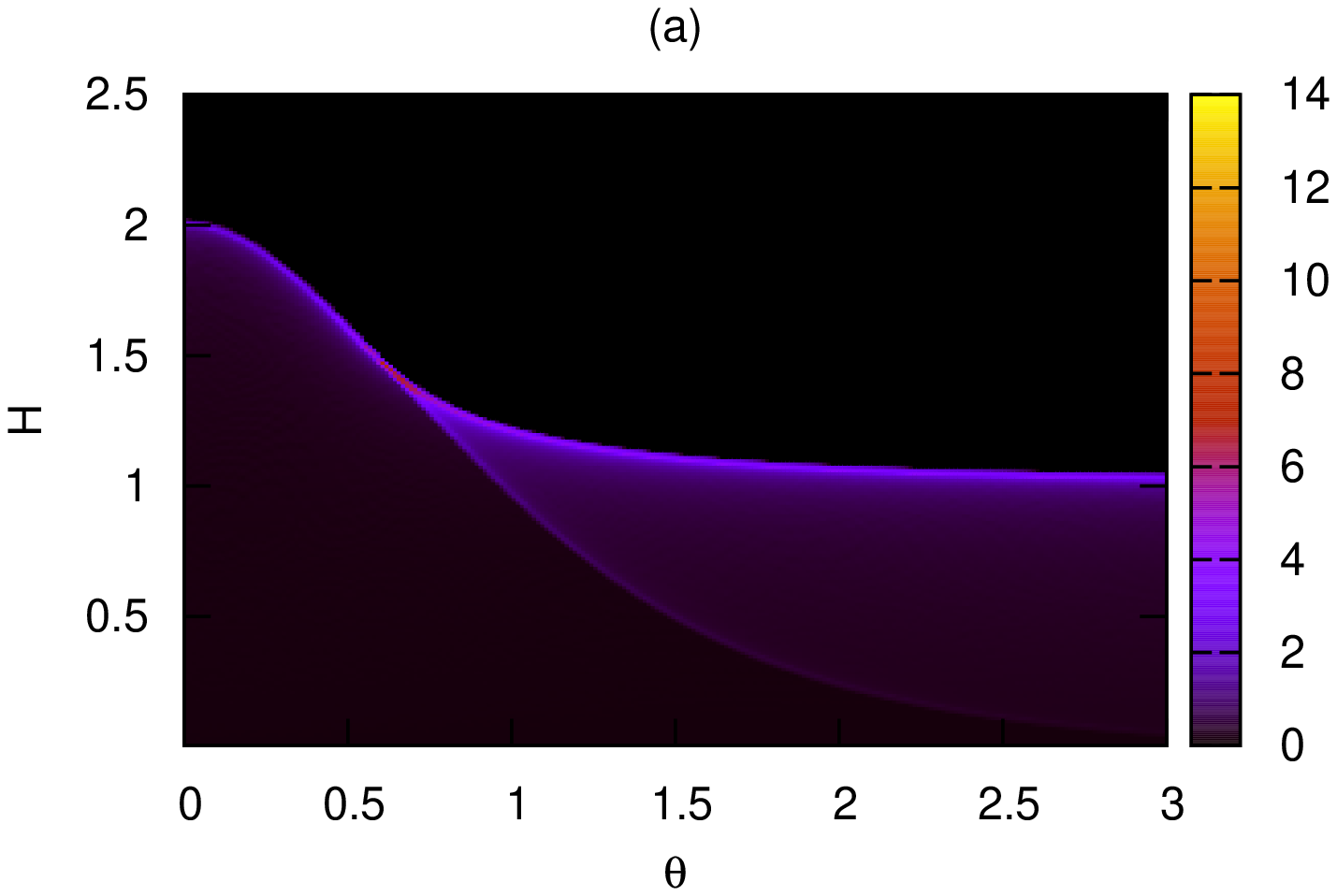}
\end{center}
\end{minipage}%
\begin{minipage}{0.5\linewidth}
\begin{center}
\includegraphics[width=1\columnwidth]{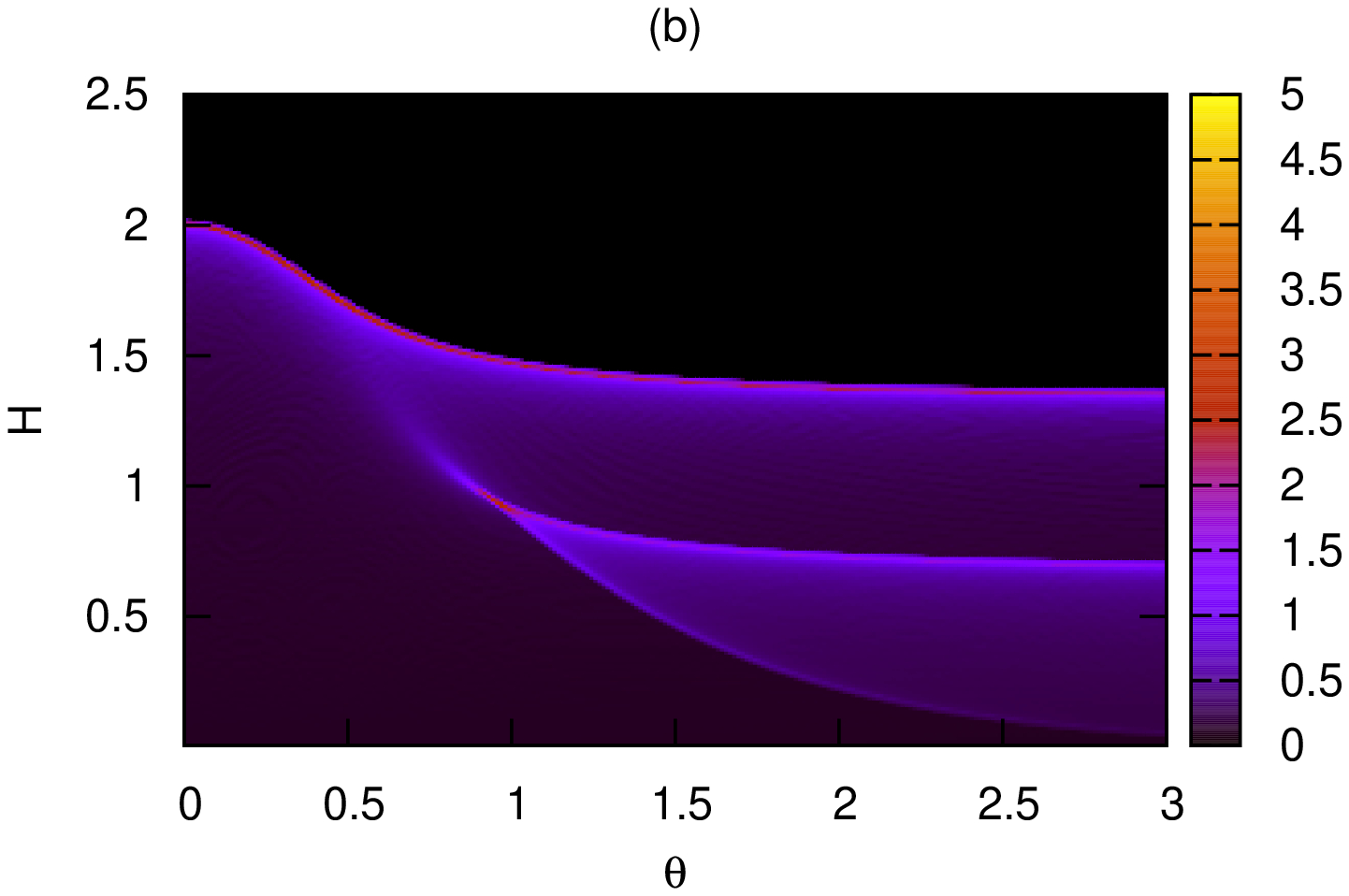}
\end{center}
\end{minipage}%
\caption{(Color online) Magnetic susceptibility on the plane $(H,\theta)$ at $T=0$ for (a) $M=3$ and (b) $M=4$.}
\label{figure1}
\end{figure}

In the case of three-spin ($M=3$) interaction Hamiltonian (\ref{HM3}), one has three different phases \cite{ZVYAGIN,FRAHM2,ZVYAGINstg,THIAGO}. There is a ferromagnetic phase with gapped excitations where the entropy vanishes and the magnetization saturates and two gapless phases which are the commensurate one for small $\theta$ and the incommensurate one for large $\theta$ \cite{ZVYAGIN,FRAHM2,ZVYAGINstg}. The ground state phase diagram can be read from the magnetic susceptibility at zero temperature, see Figure \ref{figure1}a.  In the limits $\theta=0$ and $\theta\rightarrow \infty$, we have a single chain of length $2L$ and two non-interacting chains of length $L$, respectively.

\begin{figure}[tb!]
\begin{minipage}{0.5\linewidth}
\begin{center}
\includegraphics[width=1\columnwidth]{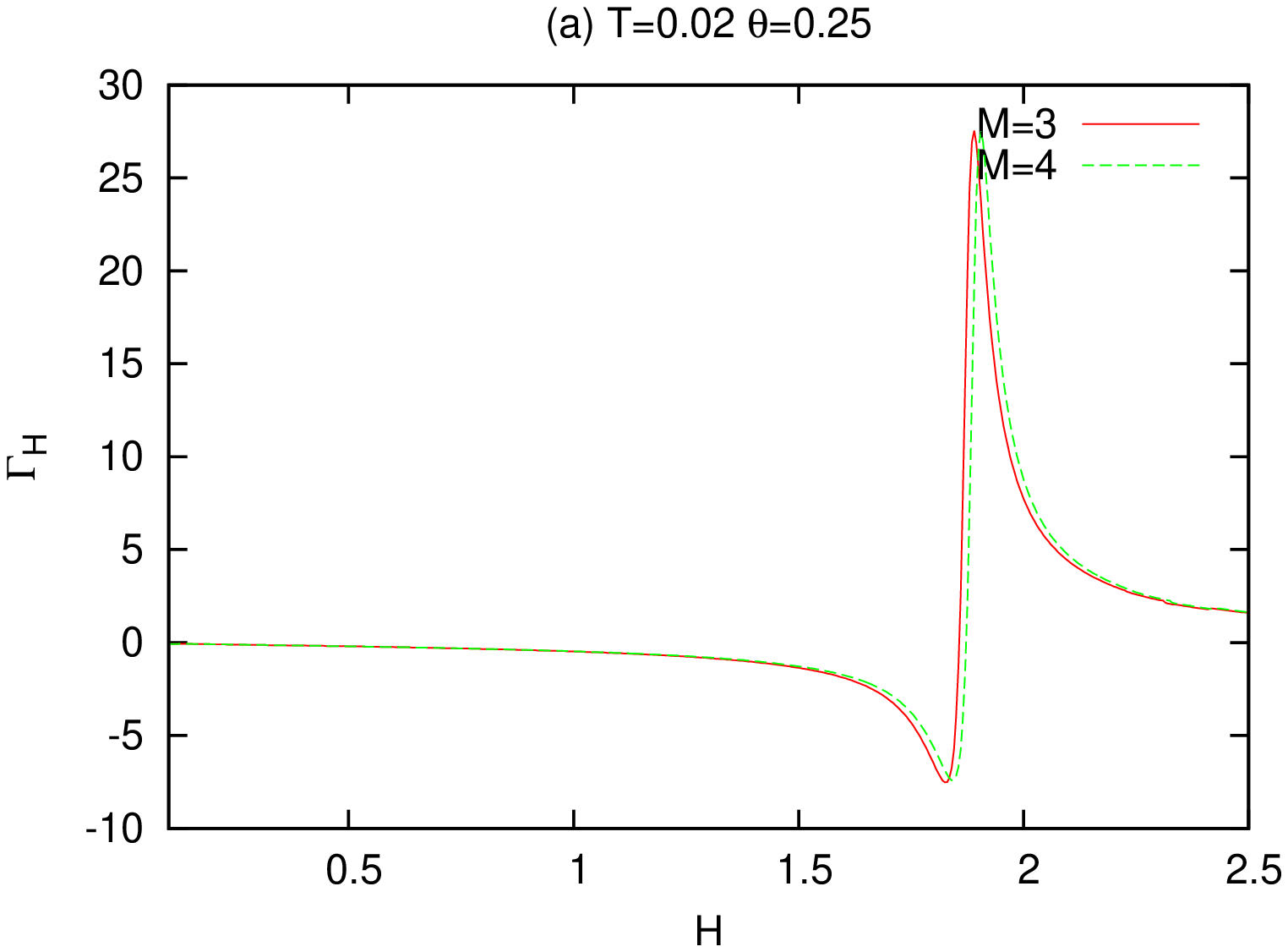}
\includegraphics[width=1\columnwidth]{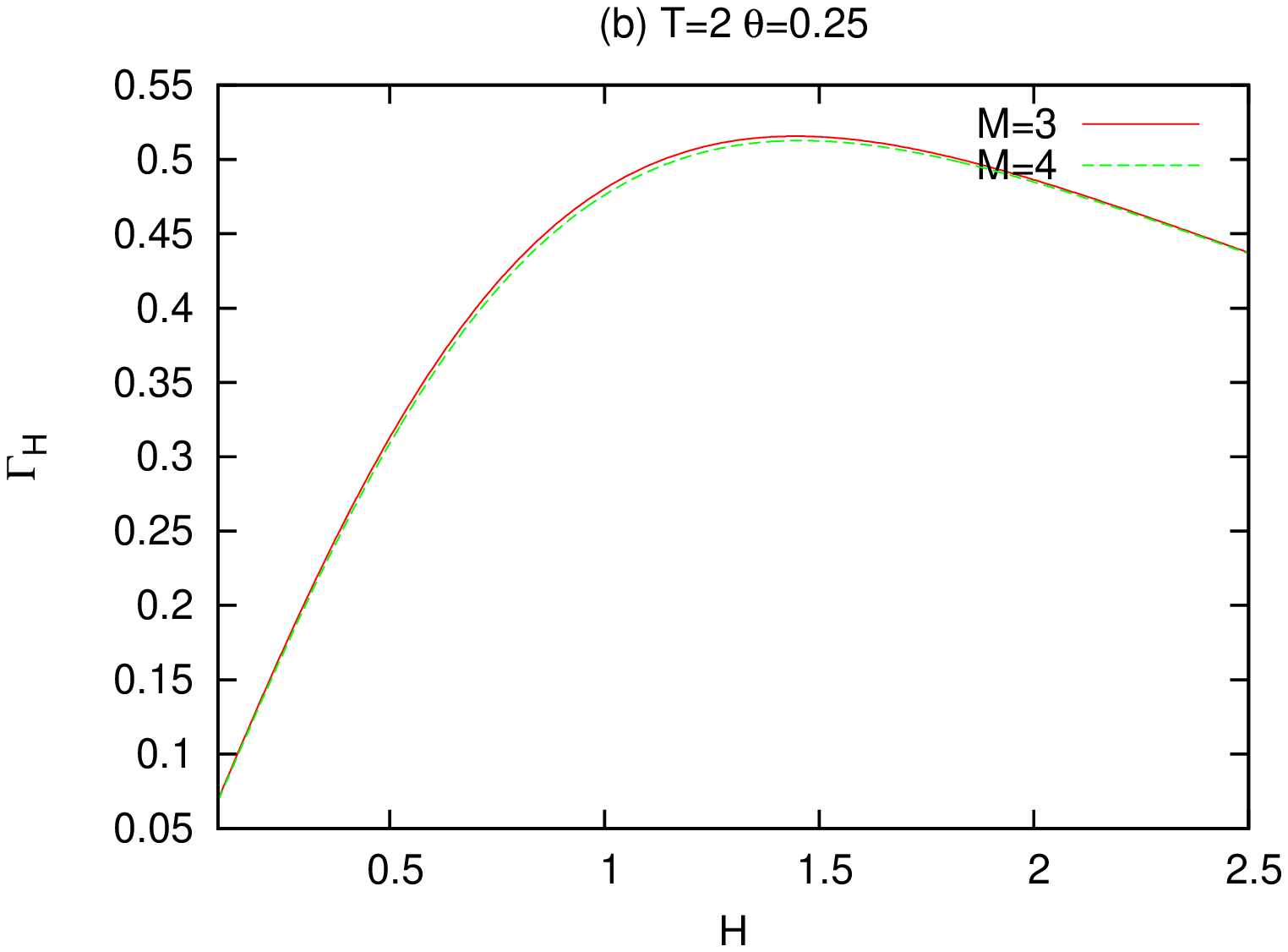}
\end{center}
\end{minipage}%
\begin{minipage}{0.5\linewidth}
\begin{center}
\includegraphics[width=1\columnwidth]{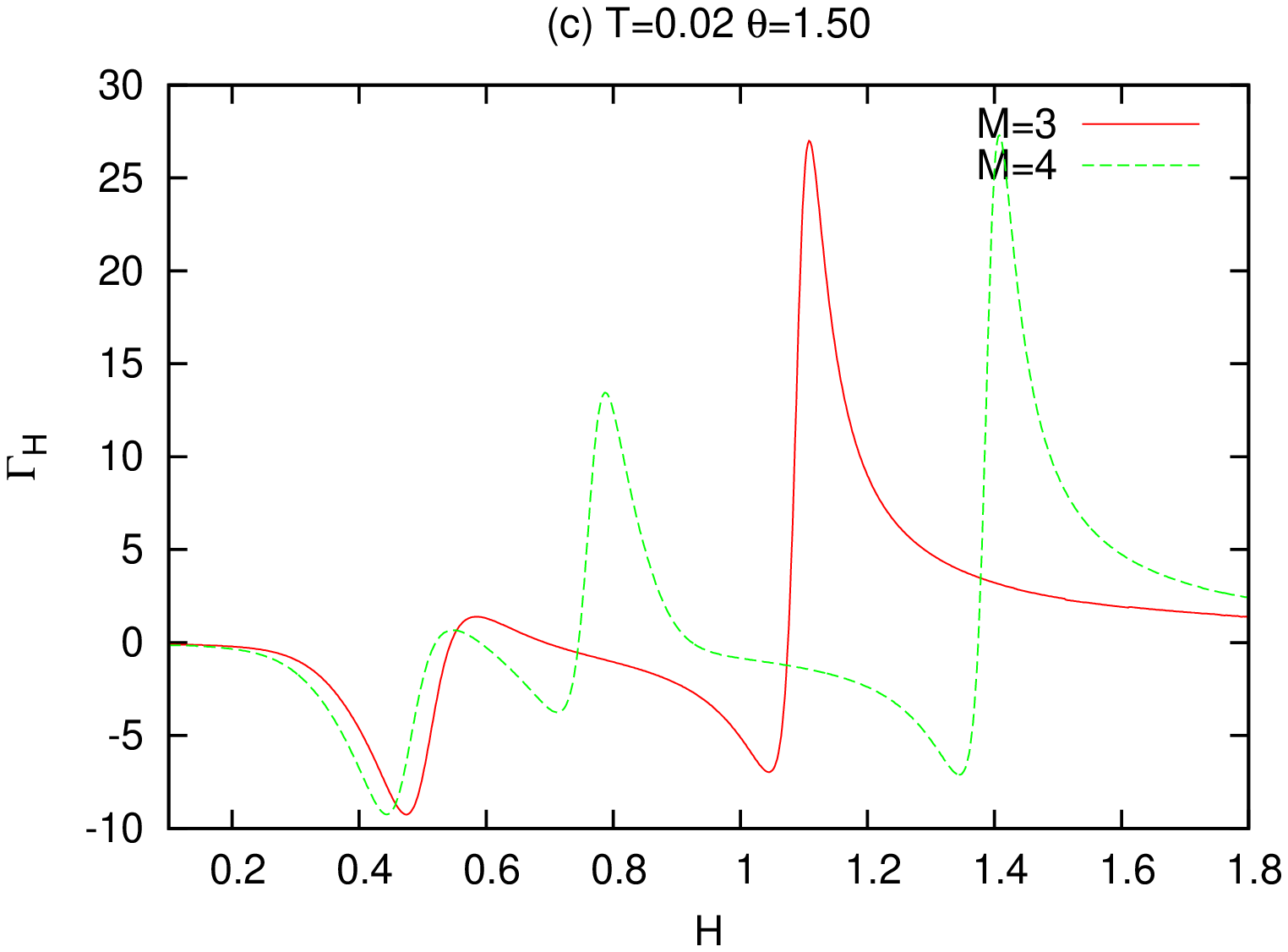}
\includegraphics[width=1\columnwidth]{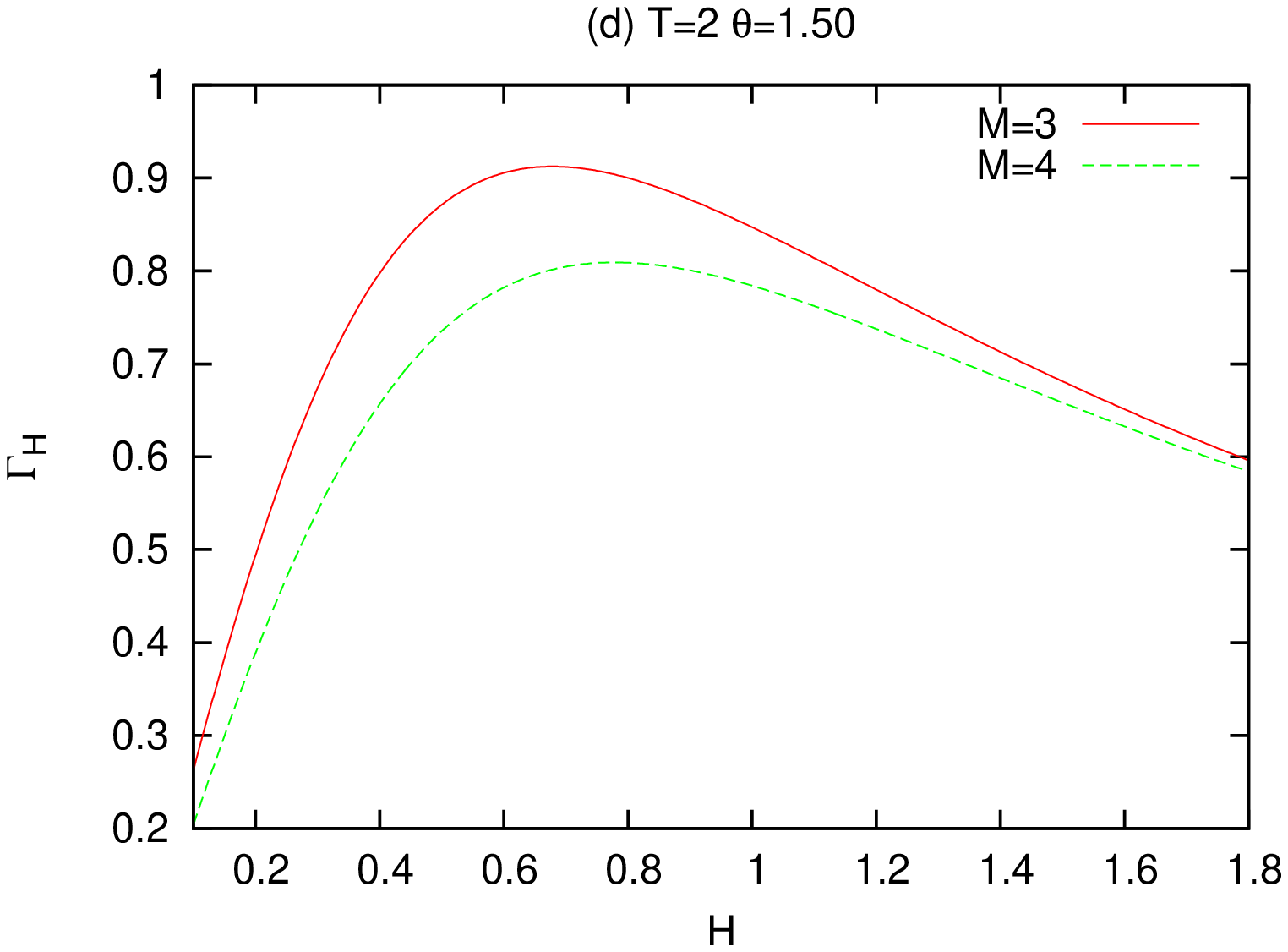}
\end{center}
\end{minipage}%
\caption{(Color online) The Gr\"uneisen parameter $\Gamma_H$ for $M=3,4$ for different values of $\theta$ and temperature: (a) $T=0.02$ and $\theta=0.25$, (b) $T=2$ and $\theta=0.25$,  (c) $T=0.02$ and $\theta=1.5$ and (d)  $T=2$ and $\theta=1.5$.
\label{figure2}}
\end{figure}

In the Figure \ref{figure2}, for the temperature $T=0.02$, we can see the transition between the gapless phases and between the gapless and ferromagnetic. This is typically signaled by sign changes of the Gr\"uneisen parameter from negative to positive values toward the higher fields values. For higher temperatures, like  $T=2$, we can see that the characteristic behaviour disappears, which implies that the thermal fluctuations are already strong enough to drive the system to excited states where no quantum phase transition effects remain.

On the other hand, in the case of four-spin ($M=4$) interaction Hamiltonian (\ref{HM4}), we have chosen $\theta_1=0$ and $\theta_2\neq 0$. This address to the case of two coupled chains of length $L$ and $2 L$. The chains have different intrachain coupling constants, which result in a superposition of phases \cite{THIAGO}. We have four different phases which can be seen in Figure \ref{figure1}b. This can also be noticed at low temperatures by the sign changes in Figure \ref{figure2}a and Figure \ref{figure2}c. In the limits $\theta\rightarrow0$ and $\theta\rightarrow \infty$, we have a single chain of length $3 L$ and two non-interacting chains of different lengths $L$ and $2 L$, respectively \cite{THIAGO}. 

\begin{figure}[th!]
\begin{minipage}{0.5\linewidth}
	\begin{center}
\includegraphics[width=\columnwidth]{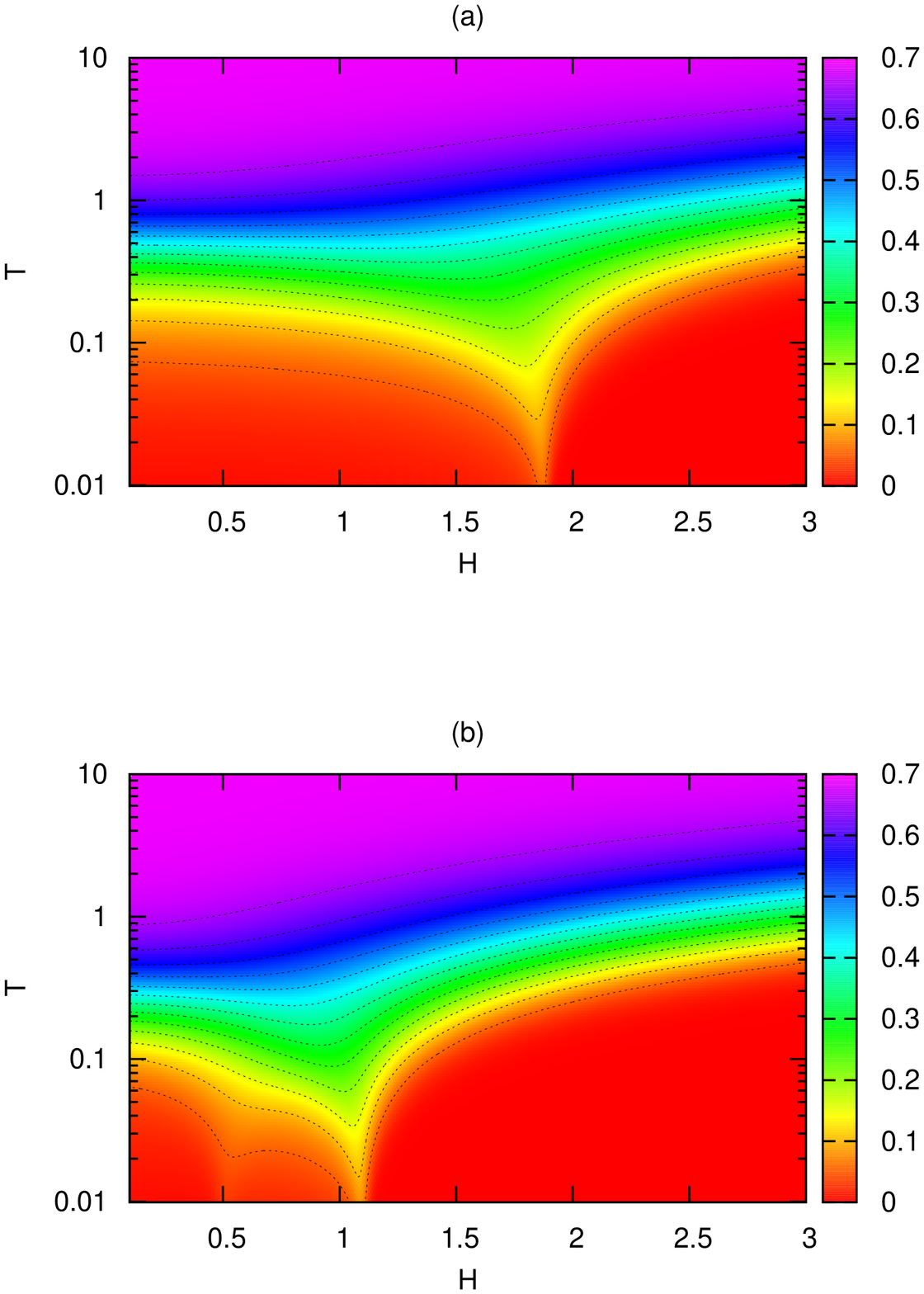}
	\end{center}
\end{minipage}%
\begin{minipage}{0.5\linewidth}
	\begin{center}
\includegraphics[width=\columnwidth]{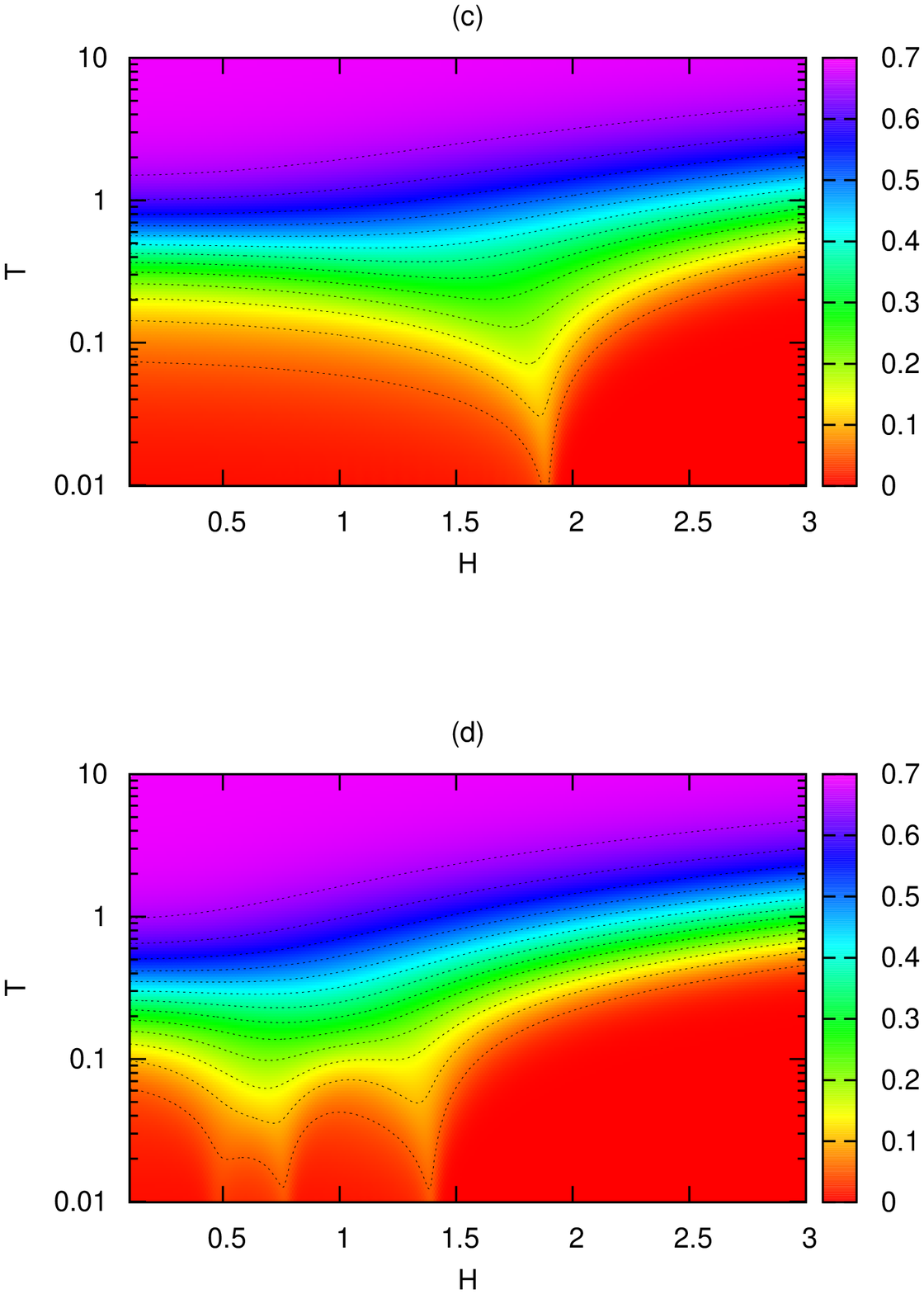}
	\end{center}
\end{minipage}%
\caption{(Color online) Entropy $S(H,T)$ for $M=3$ (a) $\theta_1=0.25$ and (b) $\theta=1.5$; and for $M=4$ ($\theta_1=0$) (c) $\theta_2=0.25$ and (d) $\theta_2=1.5$. The isentropes are for $S=0.05,0.1,0.15, \dots, 0.65$.}
\label{figure3}
\end{figure}

In Figure \ref{figure3}, we present the entropy and the isentropes for the spin chain in the $(H,T)$ plane for $M=3,4$. 
The quantum phase transitions are indicated by the isentropes, which are tilted towards the quantum critical point\cite{ROSCH2005}, showing that the entropy peaks at the critical point\cite{ROSCH2005}. The Gr\"uneisen parameter has a different sign on each side of the quantum critical point as we have shown on Figure \ref{figure2}. This is due to the fact that this parameter is proportional to the slope of the isentropes $\left(\frac{\partial T}{\partial H}\right)_S$. Moreover, a large magnetocaloric effect is indicated by the steep isentropes nearby the critical point.

\section{Conclusion}
\label{conclusion}

We studied the magnetocaloric effect in the quantum spin chains with competing interactions involving three and four spins. We exploited the tools of integrability which allow us to construct such integrable chains as well as obtain non-linear integral equations to the thermodynamic quantities. We have computed the Gr\"uneisen parameter in the thermodynamic limit as a function of magnetic field, temperature and the Hamiltonian control $\theta_i$ parameter. In addition, the entropy in the $(H,T)$ plane and its isentropes have been obtained, which indicates the existence of a large magnetocaloric effect close to the critical points.

\section*{Acknowledgments}

T.S. Tavares thanks the C.N. Yang Institute for Theoretical Physics for hospitality and FAPESP for financial support through the grant 2013/17338-4. G.A.P. Ribeiro acknowledges financial support through the grant 2012/24514-0, S\~ao Paulo Research Foundation (FAPESP).

\section*{\bf Appendix A: List of Coefficients ${\cal H}_4$}

\setcounter{equation}{0}
\renewcommand{\theequation}{A.\arabic{equation}}

In this appendix we define explicitly the Hamiltonian coefficients (\ref{HM4}).

\begin{align}
c_{0,1}&=-\frac{3 +2 \theta_2^2+2 {(\theta_2-\theta_1)}^2+ \theta_2^2 {(\theta_2-\theta_1)}^2}{6 (1+\theta_2^2) (1+{(\theta_2-\theta_1)}^2)}\\
c_{0,2}&=-\frac{3 +2 \theta_1^2+2 {(\theta_2-\theta_1)}^2+ \theta_1^2 {(\theta_2-\theta_1)}^2}{6 (1+\theta_1^2) (1+{(\theta_2-\theta_1)}^2)}\\
c_{0,3}&=-\frac{3 +2 \theta_1^2+2 \theta_2^2+ \theta_1^2 \theta_2^2}{6 (1+\theta_1^2) (1+\theta_2^2)}\\
c_{1,1}&=\frac{1}{6(1+{\theta_1}^2)(1+{\theta_2}^2)}+\frac{1}{3(1+{(\theta_2-\theta_1)}^2)}\\
c_{1,2}&=\frac{1}{6(1+{\theta_2}^2)(1+{(\theta_2-\theta_1)}^2)}+\frac{1}{3(1+{\theta_1}^2)}\\
c_{1,3}&=\frac{1}{6(1+{\theta_1}^2)(1+{(\theta_2-\theta_1)}^2)}+\frac{1}{3(1+{\theta_2}^2)}
\end{align}
\begin{align}
c_{2,1}&=\frac{\theta_1^2}{6(1+{\theta_1}^2)(1+{\theta_2}^2)}+\frac{{(\theta_2-\theta_1)}^2}{6(1+{\theta_2}^2)(1+{(\theta_2-\theta_1)}^2)}\\
c_{2,2}&=\frac{\theta_1^2}{6(1+{\theta_1}^2)(1+{(\theta_2-\theta_1)}^2)}+\frac{{\theta_2}^2}{6(1+{\theta_2}^2)(1+{(\theta_2-\theta_1)}^2)}\\
c_{2,3}&=\frac{\theta_2^2}{6(1+{\theta_1}^2)(1+{\theta_2}^2)}+\frac{{(\theta_2-\theta_1)}^2}{6(1+{\theta_1}^2)(1+{(\theta_2-\theta_1)}^2)}\\
c_{3,1}&=\frac{\theta_1}{6(1+{\theta_1}^2)(1+{\theta_2}^2)}+\frac{\theta_1-\theta_2}{6(1+{\theta_2}^2)(1+{(\theta_1-\theta_2)}^2)}\\
c_{3,2}&=\frac{-\theta_1}{6(1+{\theta_1}^2)(1+{(\theta_2-\theta_1)}^2)}+\frac{-\theta_2}{6(1+{\theta_2}^2)(1+{(\theta_1-\theta_2)}^2)}\\
c_{3,3}&=\frac{\theta_2}{6(1+{\theta_1}^2)(1+{\theta_2}^2)}+\frac{\theta_2-\theta_1}{6(1+{\theta_1}^2)(1+{(\theta_1-\theta_2)}^2)} 
\end{align}%%
\begin{align}
c_{4,1}&=\frac{{\theta_2}^2{(\theta_2-\theta_1)}^2}{6(1+{\theta_2}^2)(1+{(\theta_2-\theta_1)}^2)},  & c_{4,2}=\frac{{\theta_1}^2{(\theta_2-\theta_1)}^2}{6(1+{\theta_1}^2)(1+{(\theta_2-\theta_1)}^2)}\\
c_{4,3}&=\frac{{\theta_1}^2{\theta_2}^2}{6(1+{\theta_1}^2)(1+{\theta_2}^2)},
&c_{5,1}=\frac{{(\theta_1-\theta_2)}{\theta_2}^2}{6(1+{\theta_2}^2)(1+{(\theta_2-\theta_1)}^2)}\\
c_{5,2}&=\frac{{-\theta_1}{(\theta_2-\theta_1)}^2}{6(1+{\theta_1}^2)(1+{(\theta_2-\theta_1)}^2)},
&c_{5,3}=\frac{\theta_2 {\theta_1}^2}{6(1+{\theta_1}^2)(1+{\theta_2}^2)}\\
c_{6,1}&=\frac{-\theta_2 {(\theta_1-\theta_2)}^2}{6(1+{\theta_2}^2)(1+{(\theta_2-\theta_1)}^2)},
&c_{6,2}=\frac{(\theta_2-\theta_1){\theta_1}^2}{6(1+{\theta_1}^2)(1+{(\theta_2-\theta_1)}^2)}\\
c_{6,3}&=\frac{\theta_1{\theta_2}^2}{6(1+{\theta_1}^2)(1+{\theta_2}^2)},
&c_{7,1}=\frac{{-\theta_2}(\theta_1-\theta_2)}{6(1+{\theta_2}^2)(1+{(\theta_2-\theta_1)}^2)}\\
c_{7,2}&=\frac{-\theta_1{(\theta_2-\theta_1)}}{6(1+{\theta_1}^2)(1+{(\theta_2-\theta_1)}^2)},
&c_{7,3}=\frac{{\theta_1}{\theta_2}}{6(1+{\theta_1}^2)(1+{\theta_2}^2)}
\end{align}

\end{document}